# Beyond Retractions: Forensic Scientometrics Techniques to Identify Research Misconduct, Citation Leakage, and Funding Anomalies

Leslie D. McIntosh, PhD;  Alexandra Sinclair, PhD; Simon Linacre

## Abstract

This paper presents a forensic scientometric case study of the Pharmakon Neuroscience Research Network (PNRN), a fabricated research collective that operated primarily between 2019 and 2022 while embedding itself within legitimate scholarly publishing channels. We identified 120 research and review publications across 56 journals and 12 major publishers in which PNRN appeared as an author affiliation, funder, or acknowledged entity. These publications involved 312 distinct authors affiliated with 232 organizations across 40 countries.

To examine longitudinal author behavior, we conducted hierarchical clustering of proportional publication activity before (2015–2018), during (2019–2022), and after (2023–2025) the network's active period. Four distinct temporal publication archetypes emerged, ranging from sustained publishing across all periods to authors publishing exclusively during the network's operation. While publication timing alone cannot establish intent, these patterns reveal heterogeneous author trajectories and identify subsets whose scholarly presence is tightly coupled to the network.

Funding analyses indicate systemic risk: although few authors held grants, more than US$117 million (equivalent) in public funding was associated with individuals who participated in PNRN publications. Most PNRN-linked articles remain uncorrected, demonstrating how low-visibility misconduct persists through citation leakage. We show how forensic scientometric techniques can detect and contextualize such contamination beyond retractions.

## Background

Scientific publishing has been - and is continuing to be - manipulated to advance personal, organizational and governmental gains. Methods to achieve this include paper mills, manipulated peer review, plagiarism and a whole range of unethical activities that breach agreed research codes of conduct.

It is perhaps due to the nature of these breaches in research integrity norms, as well as the relatively recent discovery of activities such as paper mills and predatory journals, that there is a lack of research in this area compared to more established disciplines. The research that has emerged on nefarious activities tends to focus on case studies, or meta-analyses of research that when brought together paint a bleak picture of individuals and systems operating in a wholly self-interested way that benefits neither science nor society.

Research integrity issues have also become a concerning issue in recent years fuelled by the availability of effective content generating technologies using Generative Artificial Intelligence (GenAI). Yet paper mills' history can be traced to before this period. Recent advances suggest dubious and deceitful research publishing practices have been enhanced by GenAI, despite increasing awareness by funders, policymakers and authors themselves (Parker, 2024). Various means have been adopted to identify paper mill use, using trust markers (McIntosh et al., 2023) and other strategies using network-based methodologies to highlight anomalies (Cardenuto et al, 2024; Porter and McIntosh, 2024).

However, research integrity issues have not solely been a product of technological advancement, and the history of meta-research is littered with cases of peer review fraud (Dyer, 2017), so-called citation cartels and citation doping (Plevris, 2025) where again network-based analysis can uncover nefarious behaviors (Kojaku et al, 2021). The constants exhibited in these examples are the incentives in place presented by a 'publish or perish' culture and network effects where people either collaborate to provide a dishonest solution to the problem, or work together to satisfy the demands placed on them to reap the benefits of the incentives available. Yet, it should not be overlooked that the principles of open science have also opened financial opportunities to profit - both ideologically and financially (Pinto, 2020) - from newly formed publishing practices such as charging APC fees (Linacre, 2022) and cutting out library curation in journal selection.

Another aspect of research integrity is the choice to publish in journals that are known not to conduct peer review or any meaningful checks, known as predatory journals. While a key defining characteristic of these titles is the attempted deception of unaware authors (Linacre, 2022), some make the unethical choice and publish in the journals in the full knowledge that their research will not be validated (Frandsen, 2019). The impact of this author-led deception is unknown, but early research in the post-Covid era shows that the effects could be significant (Rhode et al, 2024).

## Case studies and meta-research

Individual case studies have inevitably gained a large amount of attention and notoriety, especially when using 'sting' tactics to elicit findings. This has been used numerous times to highlight predatory publishing behaviors, for example, with Bohannon (2014) and Mazieres and Kohler (2014) being the most prominent. These studies have shown not only that predatory publishers will publish anything without peer review, but also that more reputable journals will allow articles to be published without effective checks (Oviedo-García, 2024; Richardson, 2025).

Furthermore, articles that have been published with errors that should have clearly been dealt with during the submission and publishing process have also highlighted the need for more robust publisher checks to be implemented on a consistent basis. The following infamous articles clearly illustrate these problems of lack of proper oversight in the now-retracted papers: in Frontiers' Frontiers in Cell and Developmental Biology (Frontiers Editorial Office, 2024) an anatomically incorrect mouse with four testicles depicted an imaginary experiment with misspelled labels; and more recently in Springer Nature's flagship Open Access journal Scientific Reports [Jiang, 2025] which has AI-generated images, dead-linked references, and other nonsensical practices. Issues in the publishing process in these instances are clear, but other investigations have gone deeper to unearth similar and other problems.

The evolution of meta-research – or 'research about research' – has had to develop quickly due to the advances in technology, and may not be keeping pace with the rapid emergence of different research integrity challenges. Due to the scale and size of publishing outputs (over seven million articles were published in 2025 according to Dimensions from Digital Science (2025)), many meta-analyses of research integrity issues focus on specific subject areas (Aquarius, et al., 2025). Phogat et al. (2023) examined cases of misconduct in biomedical sciences and found that plagiarism and data falsification were the two main issues, but both were much more likely to be discovered than self-reported.

Earlier studies have managed to complete scoping reviews across multiple disciplines, and confirmed the findings that misconduct was more likely to be discovered than not (Xie, 2021). Another scoping review found that fabrication and falsification were the most frequent violations, accounting for almost half of all cases, followed by non-compliance with regulations, safety issues, and plagiarism (Armond et al, 2021). Medical and biological sciences dominated these results, given the scope for serious violations in the nature of the experimentation conducted.

Many other problems plague the research ecosystem, at the researcher, paper, journal, institution and governmental levels for financial, ethical, management, and even geopolitical reasons. Individual, complex issues also pose questions, such as the obscuring conflicts of interest (McIntosh & Vitale, 2023), or prevalence of hijacked journals (Albakina, 2021), as well as those now presented in this article. An editorial from the Editor-in-Chief of Science journal also pointed to the increasing tide of 'AI slop' or 'academic slop' that was threatening the scientific record, and how his journal was focused on repelling the threat it represented (Thorp, 2026). The threat is multifarious, and involves a number of ethical and integrity problems investigated in this article.

## Original Case

In early 2022, a case emerged that exemplified the vulnerabilities of the scholarly communication system to deliberate manipulation. In March, a series of tweets surfaced, raising concerns about the originality of a published research paper and bringing attention to certain discrepancies in the publication details.

An academic (Author 1) complained that their paper (Paper 1 in Journal A) was plagiarised in another published paper (Paper 2) in a different journal (Journal B). Author 1 commented that neither the editors of Journal B nor the publisher responded to requests for updates of the allegation. Author 1 did not give the title of their paper; however, they gave a link to a paper that was allegedly plagiarised (Paper 2).

This observation prompted a systematic inquiry into what became known as the Pharmakon Neuroscience Research Network (PNRN) – a fabricated research collective whose outputs spanned legitimate journals and publishers originating in more than 40 countries.

**Methods of Detection**

The first step was to look at Paper 2 through the lens of trust markers. Trust markers are key elements in publications that elicit trust (or mistrust) in an article's content. Three broad categories, each with subcategories make up trust markers - authorship, transparency and reproducibility elements (McIntosh, et al., 2023).

The first phase was to look at the individual paper to get a general idea of its structure, the authors and information (e.g., emails, institutions, network connections) then inspect the transparency trust marker elements (e.g., ethics statement, conflicts of interest, funding statements). One corresponding author used two almost identical emails associated with hotmail.com. This was unusual not because of using a non-institutional email address but because there were two emails with only minor differences - the first used a dash between names (e.g. pre-post) and the second used an underscore (e.g., pre_post). The other oddity was that the 'funder' was not listed as a GRID (https://grid.ac/) entity in Dimensions (Hook et al., 2018) of funders (see Figure 1). This could mean the funder exists but was simply not identified within GRID yet, but it warranted further inspection. Thus, a second investigative phase began.

> **Funding**
>
> The authors concede the support by the Pharmakon Neuroscience Research Network, Dhaka, Bangladesh.

**Figure 1: Funding statement from one of the PNRN publications exemplifying an unknown funding source as indicated by an absence of a hyperlink.**

## Findings: Network Composition and Activity

Using Dimensions , we conducted a free-text search within the full-text of the paper searching for the phrase "Pharmakon Neuroscience". There was no activity before 2019 (Figure 2). Between the network active years of 2019 and 2022, PNRN appeared in the funding statement, acknowledgements, or author affiliation in at least 138 publications distributed across 56 journals and 12 publishers

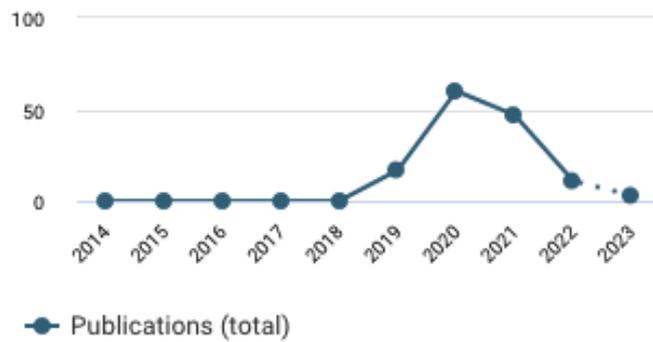

**Figure 2: Number of articles published per year by Pharmakon Neuroscience Network from 2014-2022.**

The network's visibility was amplified through social media platforms (e.g., Facebook) and ResearchGate, where calls for submissions to "special issues" under the Pharmakon banner were actively circulated (Figure 3). Co-authorship patterns revealed a recurring cluster of researchers - primarily early-career scientists - whose affiliations ranged from established universities to unverifiable institutions. Analysis of affiliated organizations indicated that while many were legitimate universities or research centers, a substantial subset were non-research entities, with a small number of commercial consultancies and unregistered institutions.

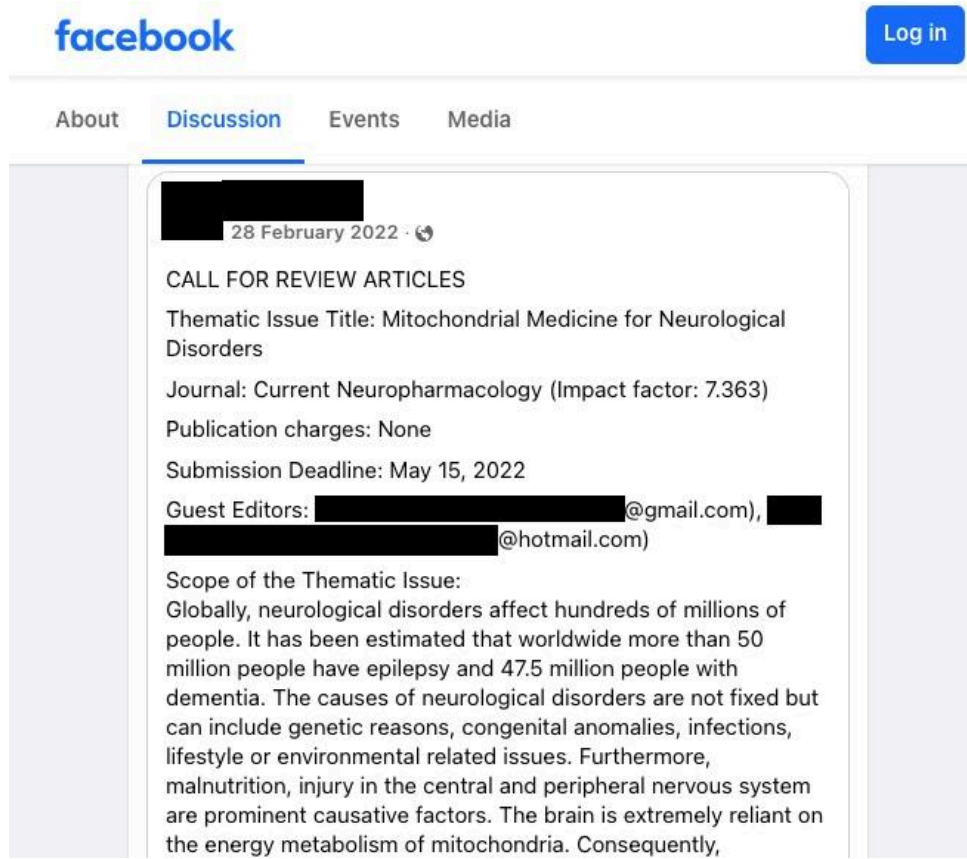

Figure 3: Example social media page for Pharmakon Special Issue Call for Review Articles

A bibliometric investigation demonstrated unusually high citation ratios (mean = 51 citations per paper) and egocentric network characteristics consistent with fabricated authorship networks (Porter & McIntosh, 2024). These features included low clustering coefficients, limited senior mentorship, and anomalously high author counts relative to field norms.

In addition to authorship anomalies, the investigation identified extensive reviewing activity. The PNRN co-authorship network alone includes 312 researchers with 2836 co-authorship links.The one author on all the publications, Md. Sahab Uddin, had a public Publons profile listed 343 verified peer reviews and 53 peer-review contributions across 10 journals, a disproportionate to his career stage as a graduate student.

Using Dimensions and Ripeta's Trust Marker framework (2023a), McIntosh and colleagues examined the PNRN corpus, which exceeded 130 papers published between 2016 and 2022, accumulating more than 6,000 citations. Citation patterns and funding acknowledgements were cross-referenced using the GRID institutional registry and metadata verification tools to assess legitimacy. One such funding statement referenced the Pharmakon Neuroscience Research Network, an entity with no corresponding record in GRID or government databases, yet appearing across numerous publications as a purported funder.

These findings raised concerns that editorial and review roles may have been leveraged to facilitate self-publication or reciprocal authorship inclusion. The network's sustained productivity - six books and over 130 articles in less than five years - was inconsistent with verifiable research activity or institutional resources.

**Institutional Response**

Upon verification of inconsistencies, the investigators (LM and SL) contacted the University of Hong Kong's research integrity office where the graduate student was enrolled. The institution responded promptly, and the researcher resigned from his program later that year. In the aftermath, only three articles affiliated with references to PNRN have been formally retracted. The first in 2022 by Springer Nature (Uddin, 2022 *Retraction Note to Exploring the Role of CLU in the Pathogenesis of Alzheimer's Disease*); the second, by Hindawi (now Wiley) *Natural Products for Neurodegeneration: Regulating Neurotrophic Signal*; and the third by Elsevier in 2025 (Kaur, et al. 2020 *Retraction notice to "Exploring the therapeutic promise of targeting HMGB1 in rheumatoid arthritis*).

As of late 2025 over, the remainder of the publications remain indexed across major bibliographic databases without retraction or editorial notice in Dimensions or Retraction Watch.

## Discussion: Systemic Implications

The Pharmakon case underscores a systemic issue rather than an isolated instance of misconduct. The PNRN operated through distributed authorship, fabricated funding entities, and exploitation of legitimate publishing channels as detailed authorship-for-sale enterprises (Porter & McIntosh, 2024). This structure enabled fraudulent outputs to gain visibility, citations, and, in some cases, to influence downstream research.

The persistence of these articles highlights the inadequacy of current retraction and metadata correction mechanisms. While initiatives such as CrossRef's open retraction database, which were greatly augmented by Retraction Watch data, have improved transparency, many problematic works remain unflagged within citation ecosystems. The concept of *citation leakage* (Linacre, 2022), the residual impact of discredited or fraudulent publications within the scholarly record, aptly characterizes this phenomenon. Each uncorrected article continues to propagate misleading citations, undermining the integrity of the research corpus.

## Research Question

After over three years since the discovery, communication, and publication of the original PNRN activities (Kincaid, 2022), an opportunity to take a retrospective look at actions taken not only from publishers, but the authors, and funders.

Primary Question: What is the research activity over time of the individuals named as authors on the original publication? This would be the publication activity of the authors and the research funding (if any).

# Methods

We inspected all papers published by the Pharmakon Neuroscience Network as an author affiliation or a 'funder'. through the following inclusion and exclusion criteria.

Using Dimensions on 23 January 2026, we used the search string "Pharmakon Neuroscience" in the full text to capture mentions of it in the author affiliations, funding statements, or acknowledgments (N=140). Retraction notices, which are notices with a separate identifier from the publisher, state that an article has been retracted and provide other information from the journal or publisher (N=2, excluded). We also limited the document type to Review Article OR Research Article OR Editorial OR Research Chapter (N=12, excluded).

Because the Dimensions database does not differentiate between authors and peer reviewers when peer reviewers are known, we also excluded papers reviewed solely by someone affiliated with PNRN (excluded N=4). In this case, there is no mention of PNRN anywhere in the paper. For the primary analyses, we also restricted papers to those with fewer than 26 authors (N=2 excluded). We will discuss these two papers separately.

From the 120 papers remaining, we extracted the standard information from Dimensions and used the following fields for analyses:

Publication ID, DOI, PMID, PMCID, Title, Abstract, Acknowledgements, Funding, S Publisher, ISSN, MeSH terms, Publication date, PubYear, Publication date (online), Publication Type, Document Type, Authors, Authors (Raw Affiliation), Corresponding Authors, Authors Affiliations, Research Organizations - standardized, GRID IDs, City of standardized research organization, State of standardized research organization, Country of standardized research organization, Funder, Funder Group, Funder Country, Grant IDs of Supporting Grants, Supporting Grants, Times cited, Recent citations, Altmetric, Fields of Research (Australian Bureau of Statistics, 2020).

The papers can be found on Figshare (McIntosh, 10.6084/m9.figshare.31224694). To better probe the data for analytics, we used Google BigQuery with assistance from DimQuery Assistant (Draux, n.d.) to generate Dimensions API queries and run the analyses, following the approaches described by Draux (2025).

We extracted all named authors from the 120 papers in our corpus, yielding 319 authors linked to unique Dimensions researcher IDs. To address authors potentially having multiple researcher IDs arising from variant name spellings or multiple institutional identifiers, we manually curated the author list, merging 319 researcher IDs into 312 distinct researcher profiles, each representing an individual author. These consolidated author profiles were then used to analyze both publication patterns (see Temporal Publication Pattern Clustering) and funding activity. For funding analyses, we examined grant start dates across multiple time periods: prior to 2019, 2019 to 2022 inclusive and 2023 to 2025 inclusive. Additionally, we calculated the number of PNRN-affiliated researchers per grant to identify collaborative funding structures within the network.

## Publication Activity using Temporal Publication Pattern Clustering

To identify distinct patterns in how researchers' publication activity changed over time, we analyzed their publication frequencies across three periods relative to PNRN activity: Before (2015-2018), During (2019-2022), and After (2022-2025).

For each researcher, we converted raw publication counts into proportions representing the relative distribution of their work across these three periods. Because these proportions are compositional data (i.e., they sum to one), we applied a centered log-ratio (CLR) transformation before analysis (Aitchison, 1982, 1986). This transformation addresses the constant-sum constraint by centering log-transformed proportions around their geometric mean, making the data suitable for standard statistical methods while preserving the underlying relationships among time periods (Pawlowsky-Glahn & Egozcue, 2011).

We then performed hierarchical cluster analysis on the CLR-transformed data using Euclidean distance, which appropriately captures the Aitchison distance in compositional space. To determine the optimal number of clusters, we used two complementary methods. First, we employed silhouette analysis (Rousseeuw, 1987) to evaluate cluster solutions from $k = 2$ to 15. The silhouette coefficient measures how well each researcher fits within their assigned cluster compared to neighboring clusters, with values ranging from -1 to +1; higher values indicate more distinct, well-separated clusters. Second, we applied the gap statistic (Tibshirani et al., 2001), which compares the within-cluster variation to what would be expected by chance, using a Monte Carlo simulation with 100 iterations. We selected the number of clusters that maximized the average silhouette width, using the gap statistic as secondary validation.

After determining the optimal cluster solution, we assigned researchers to clusters by cutting the hierarchical dendrogram at the selected k value. To aid interpretation, we calculated cluster centroids in the original proportion space, providing an average temporal publication signature for each group. All analyses were conducted in R version 4.5.2 (2025-10-31) using the heatmap and cluster packages. Complete code and data with versions are available on Figshare (10.6084/m9.figshare.31224694).

# Results

## Publications

One hundred and seventeen (N=117) research or review papers were published in 56 journals and by 12 publishers. Additionally, three book chapters were published with Springer Nature. The 120 publications averaged 51 citations, with a median of 41 and a total of 7042. Notably, the majority of publications had multiple citations, which is unusual for organically generated citations. Only five publications have not been cited, and 17 articles were cited less than 10 times. Articles appeared in established (allegedly) peer-reviewed journals rather than predatory outlets. (Table 1)

Table 1: The publisher, journal source title, number of publications, and times cited of the 120 PNRN articles.

| Publisher | Source title | Publications (n) | Times cited (n) |
|---|---|---|---|
| Bentham Science Publishers | CNS & Neurological Disorders - Drug Targets | 1 | 52 |
| | Combinatorial Chemistry & High Throughput Screening | 1 | 28 |
| | Current Drug Targets | 1 | 3 |
| | Current Gene Therapy | 4 | 74 |
| | Current Neuropharmacology | 4 | 108 |
| | Current Pharmaceutical Design | 7 | 316 |
| | Current Protein and Peptide Science | 3 | 33 |
| | Current Topics in Medicinal Chemistry | 4 | 119 |
| Bentham Science Publishers Total | | 25 | 733 |
| Elsevier | Ageing Research Reviews | 1 | 139 |
| | Biotechnology Advances | 1 | 42 |
| | Brain Research Bulletin | 1 | 102 |
| | Current Opinion in Environmental Science & Health | 1 | 0 |
| | Current Research in Pharmacology and Drug Discovery | 1 | 31 |
| | Current Research in Translational Medicine | 1 | 38 |
| | European Journal of Medicinal Chemistry | 1 | 61 |
| | European Journal of Pharmacology | 2 | 135 |
| | International Immunopharmacology | 1 | 99 |
| | Journal of the Neurological Sciences | 1 | 67 |
| | Life Sciences | 4 | 234 |
| | Pharmacological Research | 2 | 151 |
| | Phytomedicine | 2 | 188 |
| | Seminars in Cancer Biology | 2 | 172 |
| | The Science of The Total Environment | 3 | 295 |
| | Toxicology Reports | 2 | 17 |
| Elsevier Total | | 26 | 1771 |
| Frontiers | Frontiers in Cell and Developmental Biology | 3 | 236 |
| | Frontiers in Neuroscience | 1 | 39 |
| | Frontiers in Pharmacology | 5 | 153 |
| | Frontiers in Physiology | 1 | 15 |
| **Frontiers Total** | | 10 | 443 |
| Hindawi | Advances in Public Health | 1 | 3 |
| | Evidence-based Complementary and Alternative Medicine | 3 | 50 |
| | Journal of Nanomaterials | 1 | 20 |
| | Mediators of Inflammation | 1 | 8 |
| | Oxidative Medicine and Cellular Longevity | 2 | 100 |
| **Hindawi Total** | | 8 | 181 |
| IMR Press | Frontiers in Bioscience-Landmark | 1 | 2 |
| **IMR Press Total** | | 1 | 2 |
| MDPI | International Journal of Molecular Sciences | 4 | 491 |
| | Marine Drugs | 2 | 124 |
| | Molecules | 2 | 712 |
| | Pharmaceuticals | 1 | 99 |
| | Pharmacy | 1 | 28 |
| **MDPI Total** | | 10 | 1454 |
| Oxford University Press (OUP) | Journal of Pharmacy and Pharmacology | 2 | 179 |
| **Oxford University Press (OUP) Total** | | 2 | 179 |
| Royal Society of Chemistry (RSC) | Natural Product Reports | 1 | 13 |
| **Royal Society of Chemistry (RSC) Total** | | 1 | 13 |
| Springer Nature | *Book Chapters* | 0 | 54 |
| | Community Mental Health Journal | 1 | 0 |
| | Environmental Science and Pollution Research | 8 | 409 |
| | Inflammation Research | 1 | 11 |
| | Molecular Biology Reports | 1 | 11 |
| | Molecular Neurobiology | 8 | 685 |
| | Neurochemical Research | 2 | 48 |
| | Neurotoxicity Research | 2 | 108 |
| | Pharmacological Reports | 1 | 40 |
| Springer Nature Total | | 24 | 1366 |
| Taylor & Francis | Critical Reviews in Food Science and Nutrition | 1 | 107 |
| | Journal of Biomolecular Structure and Dynamics | 1 | 11 |
| Taylor & Francis Total | | 2 | 118 |
| Wiley | Archiv der Pharmazie | 1 | 120 |
| | BioMed Research International | 3 | 373 |
| | IUBMB Life | 1 | 57 |
| | Oxidative Medicine and Cellular Longevity | 2 | 89 |
| Wiley Total | | 7 | 639 |
| Wolters Kluwer | Neural Regeneration Research | 1 | 143 |
| Wolters Kluwer Total | | 1 | 143 |
| **Grand Total** | | **117** | **7042** |

## Authors

312 unique individuals were named across the publications. Twenty-nine (N=29) authors lacked persistent identifiers (e.g., ORCiD, Dimensions ID), making accountability more difficult. This could indicate that early researchers lacked the chance to establish an identity. However, as these analyses are at least three years post-publication, the individuals would most likely have had an identifier by now if they were still publishing.

In the initial investigation, one author of this publication (LDM) identified a ResearchGate page titled 'Pharmakon Neuroscience Research Network' that was associated with 38 unique profiles. The PNRN ResearchGate page was removed in 2022, along with connections from the personal profiles to PNRN. Some of the people originally associated with PNRN still have active ResearchGate profiles, though.

## Organizations

For each person listed as an author, there is one or more organizations they are associated with. If a named person of a publication is not affiliated with an institution or organization, they are typically self-classified as an independent researcher. Moreover, one person can have multiple affiliations. In PNRN, 232 organizations were linked with the 312 authors. The majority of organizations (78%) were research institutions, teaching institutions, or verifiable research companies. While many were legitimate academic or research institutions, 51 affiliations (22%) were linked to non-academic entities such as hotels, conferences, or private addresses.

## Countries

Authors have affiliations in the publications. While most authors of the research output are affiliated with a research institution, this is not always the case. Researchers also collaborate with and publish work in private institutions and occasionally as independent researchers. Based on the location provided in the publication (e.g., Arthur C. Doyle, Independent Consultant, UK), the 312 authors hail from 40 countries across six continents (Figure 4). This distributed presence shows how quickly such activity can cross borders and affect multiple national research systems.

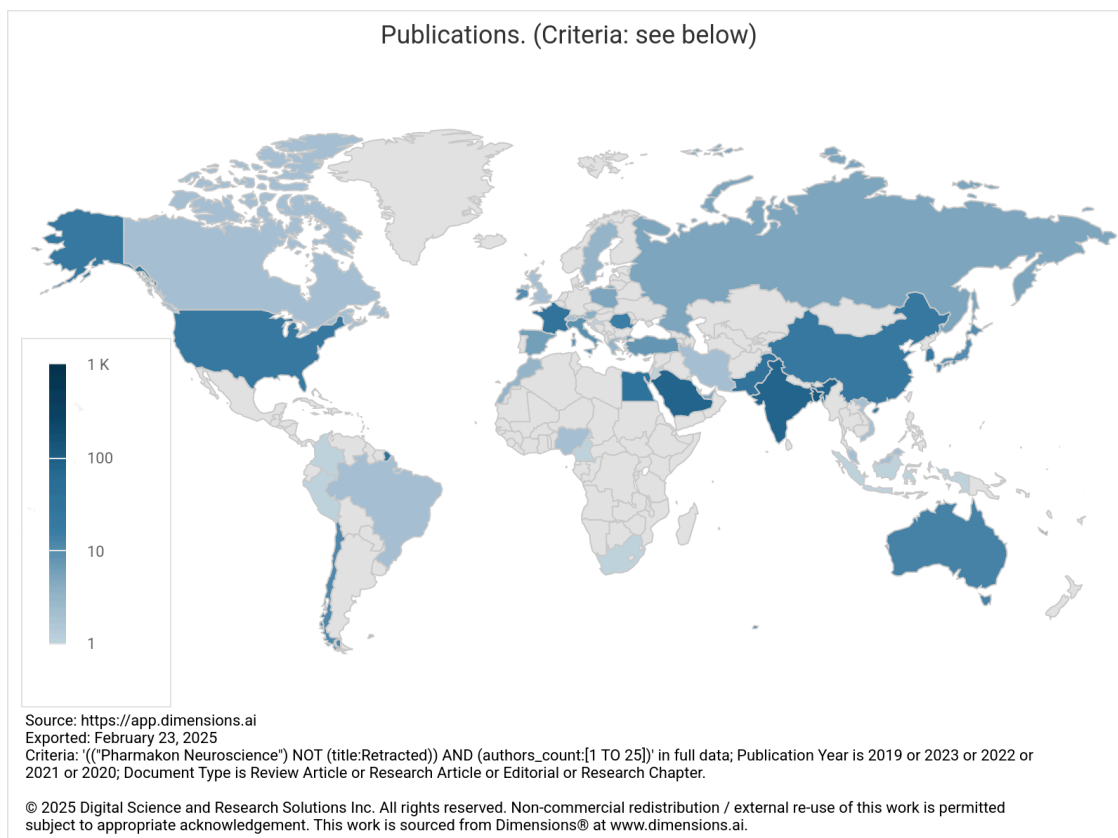

Figure 4: The country location of the 312 authors of PNRN papers.

## Multi-authored Publications

As mentioned above, two publications included more than 25 authors. One has 161 authors and the other 63. These two publications were excluded because it seems unlikely that all of the authors knew one another and actively participated in authorship-for-sale, peer-review manipulation, or other untoward activity. This does, however, raise a cautionary note about multi-authored publications, without truly knowing who else has worked on the project and who is free-riding.

# Publication Activity: Before, During, and After Participation in the Pharmakon Neuroscience Research Network

Using the clustering methods described above, we identified four general characteristics groupings of author publication activity. Clustering authors based on the proportional timing of their publications relative to the network period reveals four distinct temporal archetypes. For each cluster, the values in each column (Before, PNRN, After, Table 2) represent the average share of the author's publications that fall into each time frame, calculated across all authors in that cluster.

The groupings range from authors whose publication activity spans all phases (Cluster 1) commenced during the PNRN network period and continued at a lower volume afterwards (Cluster 2), to those authors confined entirely to the network period (Cluster 3), to those who ceased to publish after the network period (Cluster 4). Importantly, these clusters reflect patterns of publication timing rather than volume or intent, providing a structured lens through which to assess heterogeneity in author trajectories associated with the network.

Note that not all the author's publications may have been affiliated with PNRN, and we do not know if their other publications were through authorship-for-sale schemes or legitimate research. Also, Dimensions does not have funding amounts from all sources, so the total grant amount is most likely an underestimate.

Table 2: Publication practices before, during, and after PNRN resulting in four author types based on publication patterns. The table displays the mean proportional distribution of publication activity across time periods for each author cluster and the total number and percentage of authors in each cluster. Values represent the average share of each author's total output occurring before, during, and after the network period. It is important to note, however, that not all of their publications were part of PNRN.

| Cluster | Before (2015-2018) | PNRN (2019-2022) | After (2023-2025) | Authors (%) |
|---|---|---|---|---|
| 1 | 0.218 | 0.497 | 0.285 | 202 (65%) |
| 2 | 0.000 | 0.630 | 0.370 | 68 (22%) |
| 3 | 0.000 | 1.000 | 0.000 | 24 (8%) |
| 4 | 0.431 | 0.569 | 0.000 | 18 (6%) |

## Cluster 1: Publishing through Time Authors (n = 202)

This first cluster is the largest and most heterogeneous group. Authors in this cluster publish before, during, and after the network period, with activity distributed across all three windows.

On average, each author published half (49.7%) of their total publications to date during the active years of the Pharmakon Neuroscience Research Network.

This cluster captures established or ongoing contributors whose publication trajectories span multiple periods. Their presence in the PNRN window may reflect normal scholarly activity boosted by being in the network, or partial exposure rather than concentrated engagement.

## Cluster 2: Post-Network Continuation Authors (n = 68)

Authors in cluster 2 first appear during the network period and continue publishing afterward. Their activity is not confined to the PNRN window, suggesting either continued collaboration or broader publication activity beyond the period of known network operation. However, the proportion of their publications drops considerably after the PNRN window - to 37.0% of their publications.

This pattern is consistent with authors whose entry into the publication record coincides with the network period and who maintain output afterward. This group is analytically important for understanding whether network-associated activity represents a one-off event or part of a longer publication trajectory.

## Cluster 3: Network-Period-Exclusive Authors (n = 24)

The authors in cluster 3 publish exclusively during the known network period, with no observable activity before or after. This is the most temporally concentrated pattern in the dataset and represents a highly distinctive signature. While no conclusions about intent can be drawn from timing alone, this pattern aligns with short-lived or opportunistic participation and warrants careful contextual interpretation.

## Cluster 4: Pre-Network & Network-Period Authors (n = 18)

The authors in cluster 4 are active prior to the network period (43.1% of publications) and during it (56.9% of publications), but cease publishing afterward. Their output suggests disengagement or interruption following the PNRN window.

This pattern may reflect authors whose participation in the scholarly record (e.g., retirement, graduation) ends around the close of the network period. It could also indicate a change in identity in order to continue publishing without association to an alleged authorship-for-sale network. Understanding whether these authors reflect career changes, institutional factors, or bibliometric alterations requires additional contextual data.

# Funding

In addition to understanding the publication timeframe, we were interested in understanding the legitimate funding of individuals in this network. [Need to describe that an author with funding from an established granting agencies institution may not acknowledge that funding on another paper. So they could be listed on the PNRN paper and have other work not related to this network. There could also be cases where an author did not know they were on the paper.]

## Funding from Established Sources

Of the 312 researchers, 30 have participated in funded research in their careers. The grants totalled over US$118 million (equivalent) across 13 countries. This indicates that taxpayer funds have been used to support researchers whose publications exhibit questionable research practices.

Evidence that some authors later secured legitimate funding, raising questions about how fraudulent participation can be leveraged in systems of recognition and mobility. There are nine (n=9) individuals who had no grants before participating in the PNRN but who are active on grants during or after participating in PNRN. They represent seven funding bodies in seven countries (Table 3) with over US$3.1 million equivalent.

Table 3: List of funding agencies and countries of the nine individuals who did not have grants prior to publications within the PNRN network.

| Funding Agency | Country |
| --- | --- |
| National Natural Science Foundation of China | China |
| Agence Nationale de la Recherche | France |
| Croatian Science Foundation | Croatia |
| Science Foundation Ireland | Ireland |
| Sapienza University of Rome | Italy |
| Fundação para a Ciência e Tecnologia | Portugal |
| Russian Science Foundation | Russia |

# Discussion

We have presented thoughts on an unscrupulous authorship network that could have stemmed from many motivations - those people needing papers for career maintenance or growth, non-career related financial gains (e.g., selling of authorship), or personal gains (e.g., gaining a visa to live in a foreign country). However, motivation is difficult to ascertain.

This case could be politically or organisationally motivated. For example, some researchers buy authorship or use citation cartels due to political pressures or incentives within their fields, organisations, or countries. Academia and science are not immune from political influences, biases, and agendas. Researchers in politically sensitive fields may feel compelled to cite specific papers or authors aligned with particular ideological stances. Alternatively, researchers may feel organisational pressures to preferentially cite colleagues, collaborators, or superiors to gain favour or status within their institutions. These organisational citation doping (Plevris, 2025) reflect ulterior motivations beyond pure scholarship.

Of course, political and organisational motivations often intersect and cannot be entirely disentangled from personal motivations. The reward systems and power dynamics within academia and scientific institutions shape researchers' personal incentives and career ambitions. Untangling the complex motivations behind any alleged nefarious activities requires nuance and context. However, scrutinizing these motivations remains essential for upholding scientific integrity against distortions by political or organisational interests.

Issues of research integrity also overlap with questions of pedagogy, the commercialization of higher education and wider societal concerns around disinformation and misinformation. Instruction at the PhD and postdoctoral levels is often minimal, and explicit training is lacking in both research and publishing integrity expectations. Furthermore, as commercial considerations become greater for institutions as funding and incomes decrease for many higher education institutions, pressure will increase on faculty to focus on revenue-generating activities, potentially leading to a drop in postgraduate places and research activities, which in turn may lead to further deterioration of activities intended to improve research integrity practises. Finally, pressure from some governments on the type and mode of research done - for example, away from ethics-based or critical methodologies - may further impinge on best practise for research integrity.

Beyond questions of integrity, the Pharmakon Neuroscience Research Network case raises concerns regarding research security and institutional vulnerability. The exploitation of international collaborations and reviewer positions across jurisdictions demonstrates how fraudulent networks can compromise legitimate institutions, reputations, and funding systems. Motivations identified in related investigations include personal career advancement, ideological promotion, and geopolitical influence.

This case has implications across the research system. For researchers, there is a real vulnerability: once they have taken part in practices such as purchasing authorship, it can be difficult to step away from the network. For institutions, even a single compromised researcher can create reputational damage and complicate funding opportunities. At the government level, fraudulent research activity can be used not only for professional advancement but also for ideological or geopolitical purposes. A compromised researcher or institution is vulnerable to future attacks or coercion.

Vitally, when questionable work and activities go uncorrected, it weakens public confidence in science, which is central to trust.

## Reducing nefarious actions and cleaning up the scholarly record: Recommendations

The scholarly record needs a deep clean, which takes work. In this respect, we have four recommendations to enhance research integrity as practised and to strengthen the scholarly record.

First, we must take multiple approaches to detect dodgy activities. Plagiarism checkers failed to detect Uddin's papers that he allegedly plagiarized. Examining network activity and identifying standard publishing patterns by field of research plays an essential role in detection. Editors must undergo review just as researchers do to determine if they are upholding publishing ethics.

Second, the community must take responsibility. Each stakeholder in the scholarly ecosystem (publishers, institutions, funders, researchers) must play a part in upholding the integrity of science (McIntosh & Hudson Vitale, 2023b)

Third, publishers - as the publishers of the version of record - must prevent future nefarious acts. This may take on two teams working in parallel - one to reduce the flow and the other to address the anticipated barrage of questions from researchers. Researchers and institutions must show their work - transparency and reproducibility are key.

Fourth, support quality scholarship. Reward good practices, not just attention metrics. It takes much more time to produce high-quality, reproducible research that is transparently reported. Those who take the time to do this should reap the benefits.

## Limitations

It is important to note some limitations of the study, which in turn open up opportunities for future research. The primary limitation encountered was that the Dimensions database, chiefly used in the investigation, does not include funding from all sources, and therefore, it is possible that some additional sources were missed. However, this should not materially affect the outcome.

A further limitation concerns name disambiguation, a common issue in this type of research. While the list was manually curated, it is possible that one or more authors have multiple identities in the Dimensions database, which could undercount publication activity. Again, this should not have a significant impact on the reporting.

We also do not know if each author on each paper could vouch for the activities of the other authors. For example, a senior author could have a graduate student in charge of the paper, who then plagiarizes the work without the senior author's knowledge. It is also noted that this study has used the Dimensions database, but further work could be done using other curated databases such as Clarivate's Web of Science, Scopus, or Cabells' Journalytics.

Finally, we note in this paper that this forensic area of study suffers both from both being relatively new in attracting substantial numbers of academic researchers and from data often being hidden or obfuscated due to the nefarious nature of the activities in question. Further research on similar networks, such as PNRN, would be welcome, either in a case study format or using forensic scientometric techniques.

# Conclusion

The Pharmakon Neuroscience Research Network represents a paradigmatic example of how trust mechanisms in science can be subverted through the manipulation of authorship, peer review, and institutional affiliation. Its analysis through the emerging discipline of *forensic scientometrics* demonstrates the necessity of systematic, data-driven approaches to detect and mitigate academic fraud. While the network's outputs have been partially addressed through retractions and institutional interventions, the broader contamination of the scholarly record persists, underscoring the urgent need for coordinated global responses and an integrated research integrity infrastructure.

Publishers have been criticised many times over the years for inaction regarding academic fraud, despite the large number of retractions in the RetractionWatch database. While it takes time for publishers to understand and investigate potential problems, the perception of inaction further undermines trust in the academic record and in scientists themselves. Very little has been reported on the roles and responsibilities of institutions and funders.

In this case study, we have highlighted the problems created by one deceptive 'research' network and the impact this has had on the scientific record. This was not just one person. They were part of a network of authors citing one another and acknowledging a fake funder.

This case study is not just about one rogue network but how such authors can contaminate the academic record, sullying the body of knowledge for other researchers and the wider public for years to come. And yet, the study also shows the tools are there to identify such nefarious behaviour and root out the problem. By working together, using technology, and demonstrating a shared commitment, publishers, universities, funders, and researchers can clean the record and power a new wave of research and publication integrity.

## Conflict of Interest

LDM and AS are employees of Digital Science; SL is a former employee of Digital Science and current employee of Cabells. LDM is acting as VeriMe's Pre-Launch Advisors on a volunteer, unpaid basis to provide subject matter expertise, context, perspective, advice, and access to networks related to their products and services.

## CRediT statement

Leslie D. McIntosh, Digital Science, UK (corresponding author) - Conceptualization, Methodology, Formal Analysis, Writing - Original Draft, Investigation

Simon Linacre, Cabells, US - Investigation, Writing - Review & Editing

Alex Sinclair, Digital Science, UK - Software, Validation, Data Curation

## Data Availability Statement

Shareable data and code can be found at fishare https://doi.org/10.6084/m9.figshare.31224694 (Private link https://figshare.com/s/e8402dd82d8588986d95)

## Acknowledgement

Generative AI was used in accordance to the publisher guidelines. For original coding by the author LDM and data exploration, ChatGPT was used. This code was independently recreated by AS, who did not use an LLM. ChatGPT and Claude were used to help make the statistical analyses more readable, but ultimately more boring than originally planned. Generative AI (ChatGPT and Writefull's app) was also used to suggest paper titles and refine the abstract into a more concise version.

# References

Albakina, A. (2021) Detecting a network of hijacked journals by its archive. *Scientometrics* 126, 7123–7148 (2021). https://doi.org/10.1007/s11192-021-04056-0

Armond, A.C.V., Gordijn, B., Lewis, J. *et al.* (2021) A scoping review of the literature featuring research ethics and research integrity cases. *BMC Med Ethics* 22, 50. https://doi.org/10.1186/s12910-021-00620-8

Aquarius R, van de Voort M, Boogaarts HD, Reesink PM, Wever KE (2025) High prevalence of articles with image-related problems in animal studies of subarachnoid hemorrhage and low rates of correction by publishers. *PLoS Biol 23* (10): e3003438. https://doi.org/10.1371/journal.pbio.3003438

Australian Bureau of Statistics. (2020). *Australian and New Zealand Standard Research Classification (ANZSRC)*. ABS. https://www.abs.gov.au/statistics/classifications/australian-and-new-zealand-standard-research-classification-anzsrc/latest-release.

Bohannon, J. , (2013) Who's Afraid of Peer Review?.*Science* **342**,60-65. https://doi.org/10.1126/science.342.6154.60

Brown, J. (2019, March 1). *The Fake Sex Doctor Who Conned the Media Into Publicizing His Bizarre Research on Suicide, Butt-Fisting, and Bestiality.* https://gizmodo.com/the-fake-sex-doctor-who-conned-the-media-into-publicizi-1833111205

Cardenuto, J.P., Moreira, D. and Rocha, A. (2024) Unveiling scientific articles from paper mills with provenance analysis, PLoS, October 30th, 2024. https://doi.org/10.1371/journal.pone.0312666

Dimensions, Digital Science (accessed Mon 1st December, 2025)

@Dr_Meming. (2022, March 15). https://twitter.com/Dr_Meming/status/1503751505526022147

Draux, H. (n.d.). DimQuery Assistant [Custom GPT]. ChatGPT. Retrieved August 2025, from https://chatgpt.com/g/g-68000467086c8191bda31d725b124185-dimquery-assistant

Draux, H. (2025). How generative AI supports research. Research Musings. https://researchmusings.substack.com/p/how-generative-ai-supports-research

Dyer, O., (2017) Hundreds of Chinese researchers are sanctioned after mass retraction, *BMJ* 2017; 358 https://doi.org/10.1136/bmj.j3838

*Retraction notice to "Exploring the therapeutic promise of targeting HMGB1 in rheumatoid arthritis"*

[Life Sci. 258 (2020) 118164]  https://doi.org/10.1016/j.lfs.2025.123838

Fister, I., Fister, I., & Perc, M. (2016). Toward the Discovery of Citation Cartels in Citation Networks. *Frontiers in Physics, 4*, 49. https://doi.org/10.3389/fphy.2016.00049

Foster, E. M., Dangla-Valls, A., Lovestone, S., Ribe, E. M., & Buckley, N. J. (2019). Clustering in Alzheimer's Disease: Mechanisms, Genetics, and Lessons From Other Pathologies. *Frontiers in Neuroscience, 13*, 164. https://doi.org/10.3389/fnins.2019.00164

Franck, G. (1999). Scientific Communication--A Vanity Fair? *Science, 286*(5437), 53–55. https://doi.org/10.1126/science.286.5437.53

Frandsen, T.F. (2019), Why do researchers decide to publish in questionable journals? A review of the literature. Learned Publishing, 32: 57-62. https://doi.org/10.1002/leap.1214

Frontiers Editorial Office (2024) Retraction: Cellular functions of spermatogonial stem cells in relation to JAK/STAT signaling pathway. *Front. Cell Dev. Biol.* 12:1386861. https://doi.org/10.3389/fcell.2024.1386861

Hook, D. W., Porter, S. J., & Herzog, C. (2018). Dimensions: Building Context for Search and Evaluation. *Frontiers in Research Metrics and Analytics, 3*, 23. https://doi.org/10.3389/frma.2018.00023

Hoppeler, H. (2013). The San Francisco Declaration on Research Assessment. *Journal of Experimental Biology, 216*(12), 2163–2164. https://doi.org/10.1242/jeb.090779

Jiang, S. Retraction Note: Bridging the gap: explainable ai for autism diagnosis and parental support with TabPFNMix and SHAP. *Sci Rep* **15**, 43184 (2025). https://doi.org/10.1038/s41598-025-31337-y

Kaur, I., Behl, T., Bungau, S., Kumar, A., Mehta, V., Setia, D., ... & Arora, S. (2020). RETRACTED: Exploring the therapeutic promise of targeting HMGB1 in rheumatoid arthritis. *Life Sciences, 258*, 118164. https://doi.org/10.1016/j.lfs.2025.123838

Kincaid, E. (2022). How a Tweet Sparked an Investigation that Led to a PhD Student Leaving His Program. *Retraction WatchI*. https://retractionwatch.com/2022/08/24/how-a-tweet-sparked-an-investigation-that-led-to-a-phd-student-leaving-his-program/

Kojaku, S., Livan, G. & Masuda, N. (2021) Detecting anomalous citation groups in journal networks. *Sci Rep* 11, 14524. https://doi.org/10.1038/s41598-021-93572-3

Linacre, S. (2022). *Predator Effect*. https://doi.org/10.3998/mpub.12739277

Linacre, S., Rhode, S. and Berryman, K. (2024) Hidden Dangers: COVID-19-Based Research in Predatory Journals, Journal of Scholarly Publishing 2024 55:2, 165-186 https://doi.org/10.3138/jsp-2023-0063

Mazières, David; Kohler, Eddie. (2024) "Get me off Your Fucking Mailing List" (PDF). Stanford Secure Computer Systems Group, Stanford University. Retrieved November 22, 2014.

McIntosh, L. D., Whittam, R., Porter, S., Vitale, C. H., & Kidambi, M. (2023). Dimensions Research Integrity White Paper. *Figshare*. https://doi.org/10.6084/m9.figshare.21997385.v2

McIntosh, L. D., & Hudson Vitale, C. (2023a). Safeguarding scientific integrity: A case study in examining manipulation in the peer review process. *Accountability in Research*, *32*(3), 195–213. https://doi.org/10.1080/08989621.2023.2292043

McIntosh, L. D., & Vitale, C. H. (2023b). Coordinating culture change across the research landscape. *Frontiers in Research Metrics and Analytics*, *8*, 1134082. https://doi.org/10.3389/frma.2023.1134082

*Open Researcher and Contributor ID (ORCID)*. (n.d.). Retrieved September 26, 2023, from https://orcid.org/

Oviedo-García, M. Á. (2024). The review mills, not just (self-) plagiarism in review reports, but a step further. Scientometrics, 129(9), 5805-5813. https://doi.org/10.1007/s11192-024-05125-w

Parker, L., Boughton, S., Bero, L. and Byrne, J.A. (2024) Paper mill challenges: past, present, and future, Journal of Clinical Epidemiology, Volume 176, December 2024, 111549. https://doi.org/10.1016/j.jclinepi.2024.111549

Phogat, R., Manjunath, B.C., Sabbarwal, B., Bhatnagar, A.R. and Anand, D. (2023) Misconduct in biomedical research: A meta-analysis and systematic review. Journal of International Society of Preventive and Community Dentistry 13(3):p 185-193. https://doi.org/10.4103/jispcd.JISPCD_220_22

Pinto, M. F. (2020). Open Science for private Interests? How the Logic of Open Science Contributes to the Commercialization of Research. *Frontiers in Research Metrics and Analytics*, *5*, 588331. https://doi.org/10.3389/frma.2020.588331

Plevris, Vagelis (2025), From Integrity to Inflation: Ethical and Unethical Citation Practices in Academic Publishing, Journal of Academic Ethics, 23(4), https://doi.org/10.1007/s10805-025-09631-1

Porter, S. J., & McIntosh, L. D. (2024). Identifying fabricated networks within authorship-for-sale enterprises. *Scientific Reports*, *14*(1), 29569. https://doi.org/10.1038/s41598-024-71230-8

*Publons*. (n.d.). Retrieved September 26, 2023, from https://publons.com/wos-op/account/unified-auth/

R Core Team (2025). _R: A Language and Environment for Statistical Computing_. R Foundation for Statistical Computing, Vienna, Austria. <https://www.R-project.org/>

Richardson, R. A., Hong, S. S., Byrne, J. A., Stoeger, T., & Amaral, L. A. N. (2025). The entities enabling scientific fraud at scale are large, resilient, and growing rapidly. *Proceedings of the National Academy of Sciences*, *122*(32), e2420092122. https://doi.org/10.1073/pnas.2420092122

Thorpe, H. H. (2026) Resisting AI slop. *Science* 391,5-5..https://doi.org/10.1126/science.aee8267

Uddin, M.S., Kabir, M.T., Begum, M.M. *et al.* (2022). Retraction Note to: Exploring the Role of *CLU* in the Pathogenesis of Alzheimer's Disease. *Neurotox Res* **40**, 1125. https://doi.org/10.1007/s12640-022-00519-1

Xie, Y., Wang, K. & Kong, Y. (2021) Prevalence of Research Misconduct and Questionable Research Practices: A Systematic Review and Meta-Analysis. *Sci Eng Ethics* 27, 4. https://doi.org/10.1007/s11948-021-00314-9